\documentclass[12pt,a4paper]{article}

\usepackage[british]{babel}

\usepackage[a4paper,top=2cm,bottom=2cm,left=2.5cm,right=2.5cm,marginparwidth=1.75cm]{geometry}

\usepackage{amsmath, amsfonts}
\usepackage{graphicx}
\usepackage{appendix}  
\usepackage{placeins}
\usepackage{booktabs}
\usepackage{array}
\usepackage{tabularx}
\usepackage{mhchem}  
\usepackage{xcolor} 

\usepackage[colorlinks=true,
            linkcolor=blue, 
            anchorcolor=black, 
            citecolor=blue, 
            urlcolor=blue
            ]{hyperref}            

\newcommand{\DOI}[1]{doi: \href{https://doi.org/#1}{#1}}

\hypersetup{
    colorlinks=true,     
    citecolor=blue,       
    linkcolor=blue,       
    urlcolor=blue       
}

\usepackage{caption}  
\usepackage{subcaption} 
\usepackage{threeparttable} 
\usepackage{algorithm}
\usepackage{algorithmicx}
\usepackage{algpseudocode}

\usepackage{enumitem}
\usepackage{chngcntr}
\usepackage{lipsum}
\usepackage{authblk}
\usepackage[T1]{fontenc}    
\usepackage{csquotes}       
\usepackage{diagbox}

\usepackage{setspace}
\usepackage{titlesec}
\titleformat{\section} {\normalfont\Large\bfseries}{\thesection.}{1em}{}



\usepackage{lineno} 

\rightlinenumbers 

\linenumbers 

\begin{document}
\title{Irradiation Study Using QA Test Pieces of ATLAS18 ITk Strip Sensors with 80MeV Protons}

\author[1]{Y. Huang}
\author[2]{H. Li}
\author[3]{B. Crick}
\author[4]{V. Cindro}
\author[5]{A. Chisholm}
\author[6]{M. Cai}
\author[7]{H. Deng}
\author[8]{V. Fadeyev}
\author[9]{S. Hirose}
\author[7]{H. Jing}
\author[7,12]{B. Jiang}
\author[6,*]{P. Liu}

\author[1,*]{Y. Liu}
\author[6]{W. Lu}
\author[1]{H. Liu}
\author[4]{I. Mandi\'{c}}
\author[3]{R.S. Orr}
\author[6,*]{X. Shi}
\author[7]{Z. Tan}
\author[10]{Y. Unno}
\author[11]{M. Ullan}
\author[6]{S. Wang}
\author[6]{Z. Xu}
\affil[1]{\raggedright \small School of Science, Shenzhen Campus of Sun Yat-sen University, Shenzhen, Guangdong, 518107, China}
\affil[2]{\small Physics Department, Tsinghua University, Beijing 100084, China}
\affil[3]{\small Department of Physics, University of Toronto, 60 Saint George St., Toronto, Ontario M5S1A7, Canada}
\affil[4]{\small Experimental Particle Physics Department, Jožef Stefan Institute, Jamova 39, SI-1000 Ljubljana, Slovenia}
\affil[5]{\small School of Physics and Astronomy, University of Birmingham, Birmingham B152TT, United Kingdom}
\affil[6]{\small Institute of High Energy Physics (IHEP), Chinese Academy of Sciences, 19B Yuquan Road, Beijing 100049, China}
\affil[7]{\small Spallation Neutron Source Science Center, Dongguan 523803, China}
\affil[8]{\small Santa Cruz Institute for Particle Physics (SCIPP), University of California, Santa Cruz, CA 95064, USA}
\affil[9]{\small Institute of Pure and Applied Sciences, University of Tsukuba, 1-1-1 Tennodai, Tsukuba, Ibaraki 305-8571, Japan}

\affil[10]{\small Institute of Particle and Nuclear Study, High Energy Accelerator Research Organization (KEK), 1-1 Oho, Tsukuba, Ibaraki 305-0801, Japan}
\affil[11]{\small Instituto de Microelectrónica de Barcelona (IMB-CNM), CSIC, Campus UAB-Bellaterra,
08193 Barcelona, Spain}
\affil[12]{\small School of Nuclear Science and Technology, University of South China, Hengyang 421001, China}

\affil[*]{\small Corresponding author email: \texttt{peilian.liu@cern.ch (P. Liu); liuy2399@mail.sysu.edu.cn (Y. Liu); xin.shi@cern.ch (X. Shi)}}

\date{}  

\maketitle

\begin{abstract}
\noindent\indent The ATLAS experiment is planning a complete replacement of its inner detector(ID) with a new all-silicon inner tracker (ITk) for the ATLAS Inner Tracker Phase-2 upgrade. The ATLAS18 silicon strip sensors are designed to operate up to the integrated luminosity of 4000 fb$^{-1}$, which corresponds to the maximum fluence of $1.6 \times 10^{15} \, \text n_{\text{eq}} / \text{cm}^2$ (including safety factor). 
To enhance the quality assurance (QA) program to monitor the key properties of the sensors, the strip sensor community is considering to include China Spallation Neutron Source (CSNS) as a proton irradiation site and Institute of High Energy Physics (IHEP) as a QA test site. 
A total of 18 ATLAS18 ITk QA test pieces were irradiated with $6.0 \times 10^{14}$, $1.6 \times 10^{15}$, and $2.6 \times 10^{15} \, \text n_{\text{eq}} / \text{cm}^2$ protons at CSNS, and measured at IHEP, including IV (leakage current-voltage), CV (capacitance-voltage) and CCE (charge collection efficiency) measurements. 
The upgraded irradiation setup at CSNS and measurement setup at IHEP are shown in this paper. Irradiated samples were exchanged between IHEP, Ljubljana and Birmingham to cross-check CCE measurements. 
\end{abstract}

\textbf{Keywords}: Silicon strip sensors, ATLAS ITk, China Spallation Neutron Source.  


\section{Introduction}
\subsection{The upgrade of HL-LHC}
\noindent\indent The upgraded HL-LHC will start operating
at an ultimate peak instantaneous luminosity of $7.5 \times 10^{34} \, \text{cm}^{-2} \, \text{s}^{-1}$, which corresponds to a mean
of approximately 200 inelastic proton-proton collisions per beam crossing (pile-up) \cite{atlasTDR}. Over the
more than ten years of operation the ATLAS detector aims to collect data set for $4000 \, \text {fb}^{-1}$. Because of this increased luminosity and the resulting radiation damage,  the ATLAS tracking system has to withstand an integrated hadron fluence of $1.6 \times 10^{15} \, \text n_{\text{eq}} / \text{cm}^2$ in the strip region at the end of life\cite{atlasLHC}. 
The new,  fully silicon-based inner tracker is being built for this purpose. 

\subsection{Quality assurance program for the strip sensors}
\begin{figure}[ht]
\centering
\includegraphics[width=0.45\textwidth]{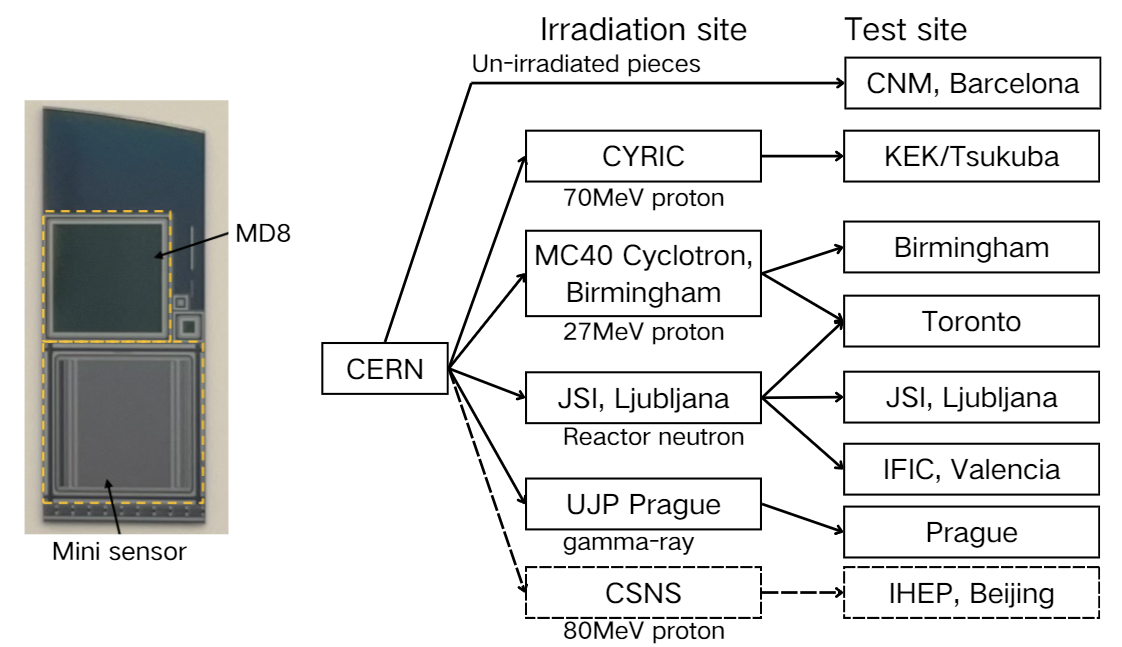}
\caption{\label{fig:1.2_MiniSensorMD8+QAFlowchart.} Photo of the QA test pieces: a mini sensor and an MD8 (left). Flow chart of QA test pieces for the QA program (right).}
\end{figure}

\noindent\indent QA is defined as the part of quality management focused on ensuring that quality requirements will be met during production \cite{Miguel2020QAmethodology}. One of the test pieces used for QA program  is shown in Figure \ref{fig:1.2_MiniSensorMD8+QAFlowchart.} (left). It contains a miniature strip sensor and a diode. The mini sensor is used for CCE measurements. It consists of 104 strips of 8 mm length, each biased through a $1.5\ \mathrm{M\Omega}$ polysilicon resistor. The mini strip sensor has the same design as the MAIN sensor but with size of $ 10 \times 10 \, \text{mm}^2 $ \cite{hara2020charge}. The other part of the structure is the monitor diode with size of $ 8 \times 8 \, \text{mm}^2 $ (MD8) for IV and CV measurements. 

\FloatBarrier

Figure \ref{fig:1.2_MiniSensorMD8+QAFlowchart.} (right) shows a flow chart of QA test pieces during the pre-production. All QA pieces were first delivered to CERN, and then distributed to each irradiation site. There are 4 irradiation sites including proton, neutron, gamma irradiation, and 7 test sites now. The strip sensor community is considering to include CSNS ( China Spallation Neutron Source) as a proton irradiation site and IHEP (Institute of High Energy Physics) as a QA test site. The QA test pieces were shipped from CERN, and irradiated at CSNS with protons, and then measured IV, CV, CCE characteristics at IHEP.

\subsection{The APEP beam line at CSNS}
\begin{figure}[h!]
    \centering
    \begin{subfigure}[b]{0.5\textwidth}
        \centering
        \includegraphics[width=\textwidth]{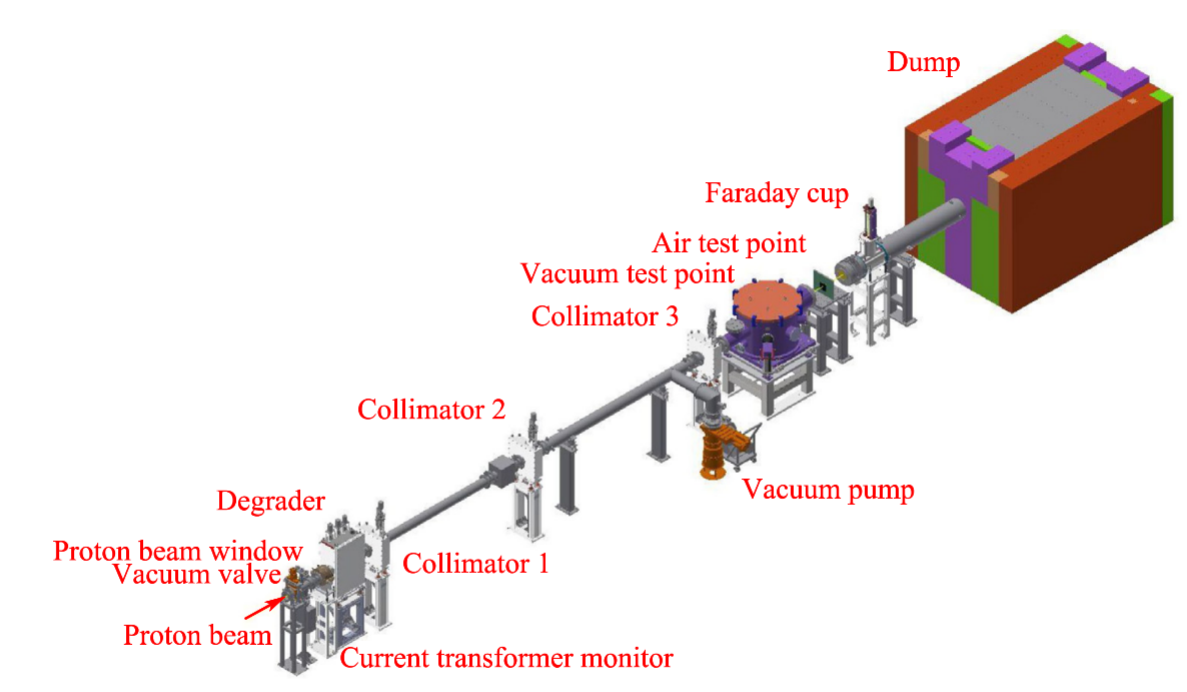}
    \end{subfigure}
    \hfill
    \begin{subfigure}[b]{0.44\textwidth}
        \centering
        \includegraphics[width=\textwidth]{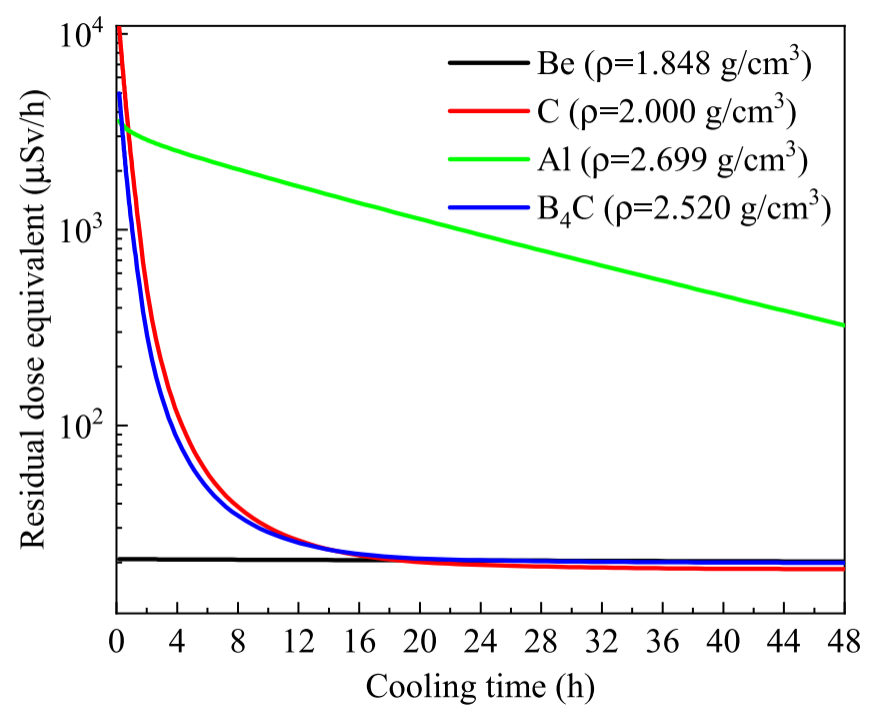}
    \end{subfigure}
    \caption{The main composition of the Associated Proton beam Experiment Platform (left). The change in the residual dose with cooling time at 30 cm from the beam center for four materials (beryllium, graphite, aluminum, boron carbide) irradiated by 0.2-\textmu A, 80-MeV proton beams for six days (right) \cite{LIU2022APEP}.}
    \label{fig:1.3_CSNS_APEP_2.1_residential_dose}
\end{figure}

\noindent\indent CSNS is a large multi-disciplinary facility for scientific research and applications. Associated Proton beam Experiment Platform (APEP) located in CSNS offers proton beam with tunable energy from 10 to 80 MeV, and a beam spot with tunable sizes from $10\, \text{mm} \times 10\, \text{mm} $ to $50\, \text{mm} \times 50\, \text{mm} $. The uniformity of beam spot is better than 90\% \cite{LIU2022APEP}.The APEP beam line is approximately 14.5 m long, as shown in Figure \ref{fig:1.3_CSNS_APEP_2.1_residential_dose} (left). There are two test points at the proton beam line, a vacuum test point and an air test point, for irradiation experiments.

\FloatBarrier

\section{Setups and measurements}
\subsection{Irradiation}
\noindent\indent 18 QA test pieces were exposed to proton beam with energy of 80 MeV, size of 20 mm $\times$ 20 mm, and beam intensity of $1.06 \times 10^{10} \, \text{protons}/\text{cm}^2/\text{s}$ (after collimation), provided by CSNS. 4 pieces were irradiated to an actual fluence of $2.6 \times 10^{15} \, \text n_{\text{eq}} / \text{cm}^2$, 12 pieces to around $1.6 \times 10^{15} \, \text n_{\text{eq}} / \text{cm}^2$, and 2 pieces to around $6.0 \times 10^{14} \, \text n_{\text{eq}} / \text{cm}^2$.

Samples were attached to the support structure inside the chamber by Kapton tape, and cooled with Peltier and water cooling to maintain the temperature at -20 \textdegree C. Dry air flowed into the chamber to limit the relative humidity during irradiation to around 5\%. 
In the previous irradiation setup \cite{LIHui2024169288}, aluminum plate was used as the support material for samples during irradiation. However, for a long-term irradiation the radionuclide \(\ce{^{22}Na}\) produced by the \(\ce{^{27}Al}(p,x)\ce{^{22}Na}\) reaction has a long half-life of 2.6 years \cite{LIWL2024}, resulting in high residual dose, posing risk to experimental personnel. Graphite is a good alternative choice, with low-activation and good thermal conductivity. Figure \ref{fig:1.3_CSNS_APEP_2.1_residential_dose} (right) shows the change in the residual dose with cooling time for four materials irradiated by 0.2\(\mu\)A, 80MeV proton beams for six days \cite{LIU2022APEP}. The residual dose of Al is much higher than that of C after irradiation.
After considering the activation and thermal conductivity, a 2 mm thick L-shape graphite plate was chosen for irradiation setup support structure.

\begin{figure}[h!]
    \centering
    \begin{subfigure}[b]{0.44\textwidth}
        \centering
        \includegraphics[width=\textwidth]{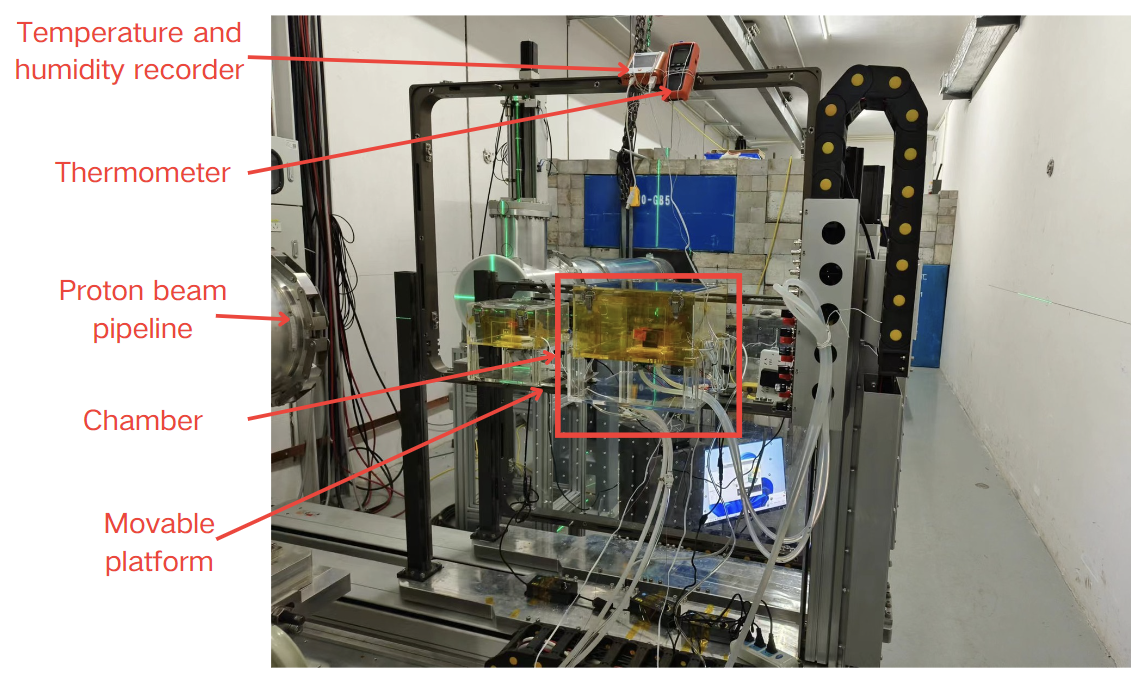}
    \end{subfigure}
    \hfill
    \begin{subfigure}[b]{0.55\textwidth}
        \centering
        \includegraphics[width=\textwidth]{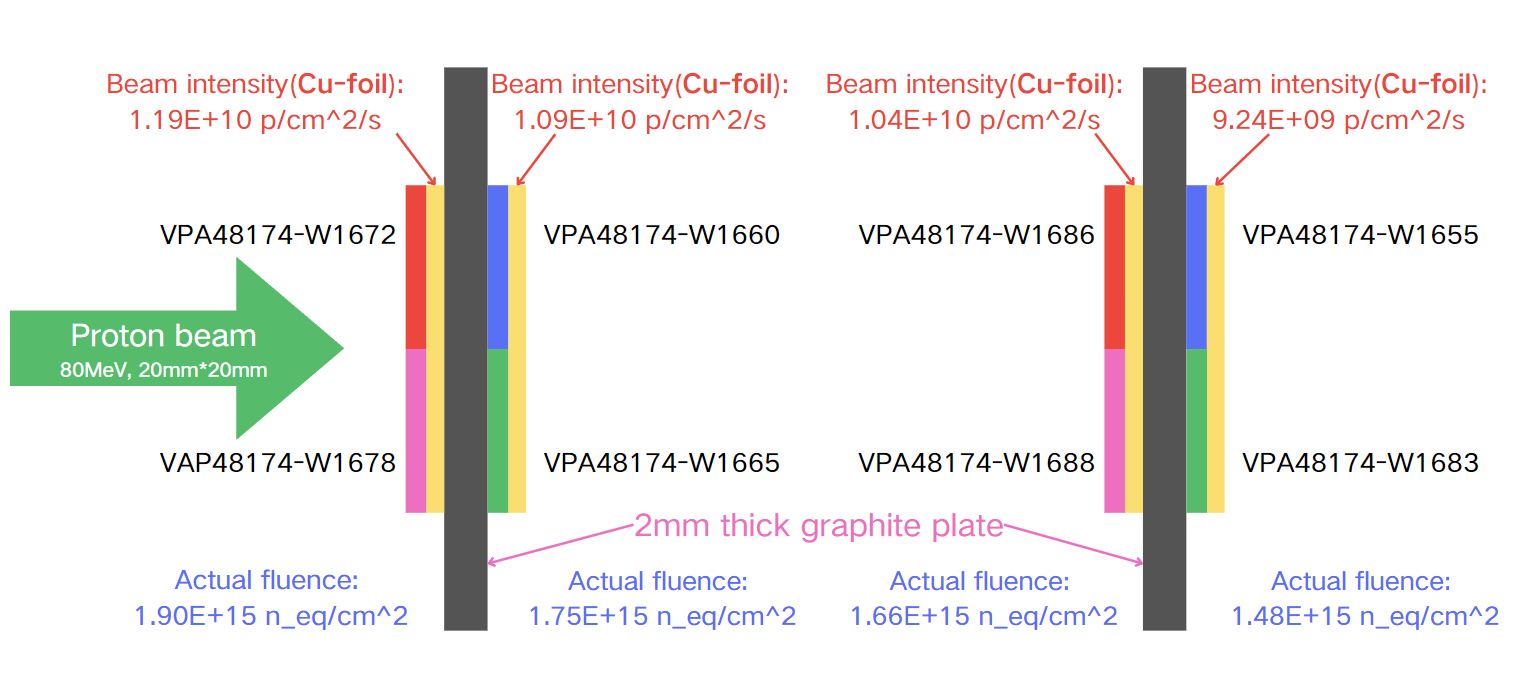}
    \end{subfigure}
    \caption{Irradiation setup placed at air test point of APEP at CSNS (left). A sketch of samples and Cu-foils in the proton beam, with estimates of the beam intensity and actual fluence (right).}
    \label{fig:2.1_irradiation_setup_beam_intensity}
\end{figure}

The irradiation setup is shown in Figure \ref{fig:2.1_irradiation_setup_beam_intensity} (left). Figure \ref{fig:2.1_irradiation_setup_beam_intensity} (right) shows samples and Cu-foils in the direction of the proton beam. The proton beam intensity was measured using Cu foils attached to the samples. The foil activation was evaluated for this purpose, with the measurement precision of around ±4.9\% \cite{LIWL2024}. The NIEL (Non-Ionizing Energy Loss) hardness factor for proton with 80 MeV energy found in the literature \cite{NIEL_2000} was 1.427, which was used for converting proton fluence to 1 MeV neutron equivalent fluence ($\text n_{\text{eq}} / \text{cm}^2$). The irradiation time could be calculated from the accelerator monitoring feedback. 

\FloatBarrier

\subsection{Thermal Annealing Procedure}
\noindent\indent Since the degree of thermal annealing significantly affects the post-irradiation leakage current of sensors, all the samples where thermally annealed for 80 minutes at  60 $^\circ\mathrm{C}$ in accordance with the
QA program \cite{hirose2023atlas}. This ensured that post-irradiation, all sensors possessed the same thermal history \cite{allport2019experimental}. 

\subsection{IV and CV measurements}
\begin{figure}[h!]
    \centering

    \begin{subfigure}[b]{0.45\textwidth}
        \centering
        \includegraphics[width=\textwidth]{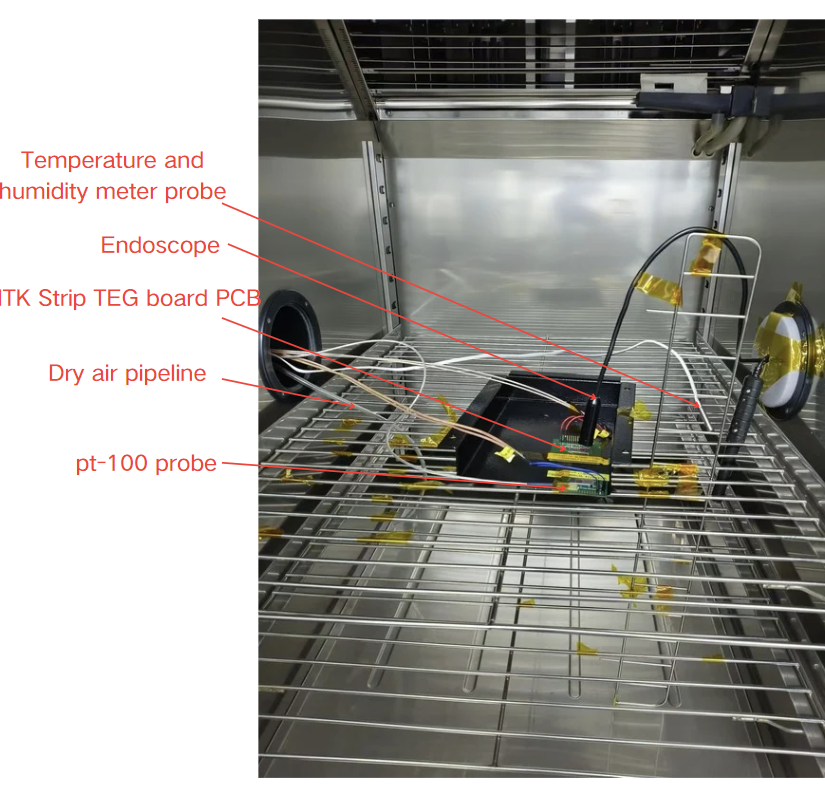}
        
    \end{subfigure}
    \hfill
    \begin{subfigure}[b]{0.46\textwidth}
        \centering
        \includegraphics[width=\textwidth]{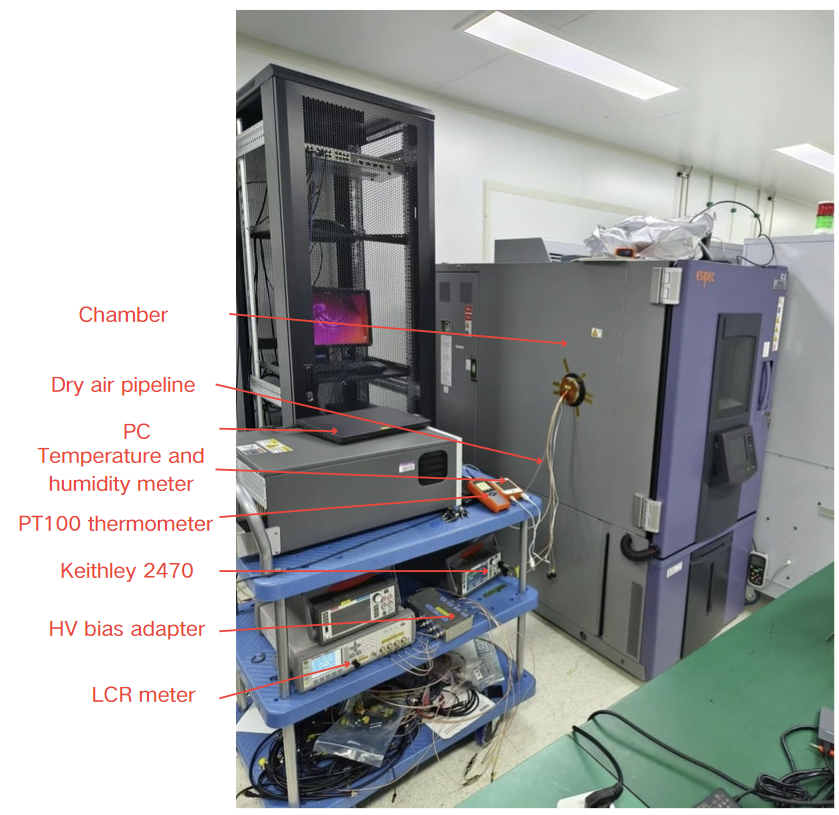}
        
    \end{subfigure}

    \caption{MD8 and PCB in the climate chamber (left). Setups for IV and CV measurements (right).}
    \label{fig:2.3_IV_CV_pcb_and_chamber}
\end{figure}

\noindent\indent To estimate the IV and CV characteristics of the sensors, the MD8 was attached to the ITK Strip TEG board PCB and placed in the chamber, as shown in the Figure \ref{fig:2.3_IV_CV_pcb_and_chamber}.  The temperature and humidity conditions for the IV and CV measurements were the same. The measurement temperature for the un-irradiated MD8 was 20 $^\circ$C, while for the irradiated MD8 it was -20 $^\circ$C. The relative humidity was below 10\% in both cases. 

\FloatBarrier

The LCR meter (Keysight E4980A) for CV measurements is an auto-balancing bridge instrument, which has an internal DC bias function for applying bias voltage to device-under-tests (DUT)s. The internal bias source can typically output a variable bias voltage of up to \( \pm 40 \, \mathrm{V} \) \cite{LCRe4980a_handbook}.
During the measurement, a reverse bias voltage of up to \( 700 \, \mathrm{V} \) was applied to the MD8 backplane while the bias ring was kept at ground. Therefore, an external DC voltage source (Keithley 2470) and a HV bias adapter were required to apply DC bias voltage that exceeds the limits of the internal DC bias function. The HV bias adapter, as a protection circuit, was added to isolate high voltage from the LCR meter, which was voltage sensitive and cannot withstand DC voltages on the order of \( 1 \, \mathrm{kV} \) \cite{HV_capacitance2009}.
An open circuit calibration ensured appropriate accuracy. Impedance measurements at \( 1 \, \mathrm{kHz} \) with CR in parallel and an AC amplitude of \( 1 \, \mathrm{V} \) were applied to extract the capacitance values \cite{ullan2019quality}.

\subsection{CCE Measurements}
\begin{figure}[h!]
    \centering

    \begin{subfigure}[b]{0.44\textwidth}
        \centering
        \includegraphics[width=\textwidth]{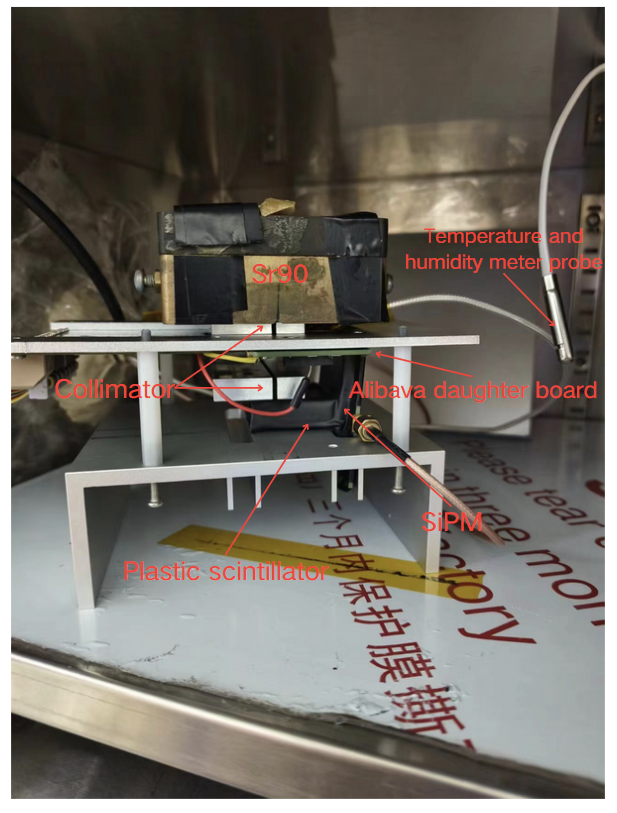}
        
    \end{subfigure}
    \hfill
    \begin{subfigure}[b]{0.4\textwidth}
        \centering
        \includegraphics[width=\textwidth]{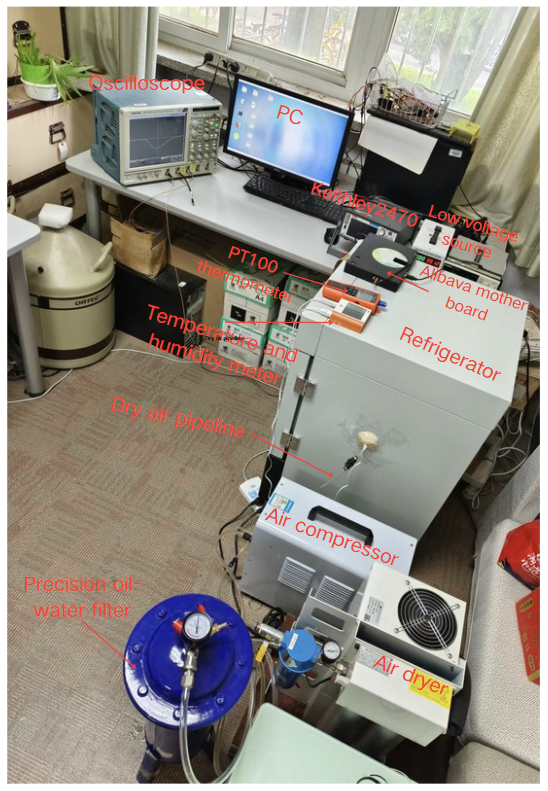}
        
    \end{subfigure}

    \caption{Front-end CCE measurement boards inside the refrigerator (left). CCE measurement setup (right).}
    \label{fig:2.4_CCE_setup}
\end{figure}

\noindent\indent The sample for the CCE measurements was the mini sensor. The CCE measurement setup is shown in Figure \ref{fig:2.4_CCE_setup}, both irradiated and un-irradiated mini sensors were measured at $-20\,^\circ\mathrm{C}$. Dry air flowed into the refrigerator to maintain the humidity below 10\% RH.

\FloatBarrier

\begin{figure}[!htb]
\centering
\includegraphics[width=0.6\textwidth]{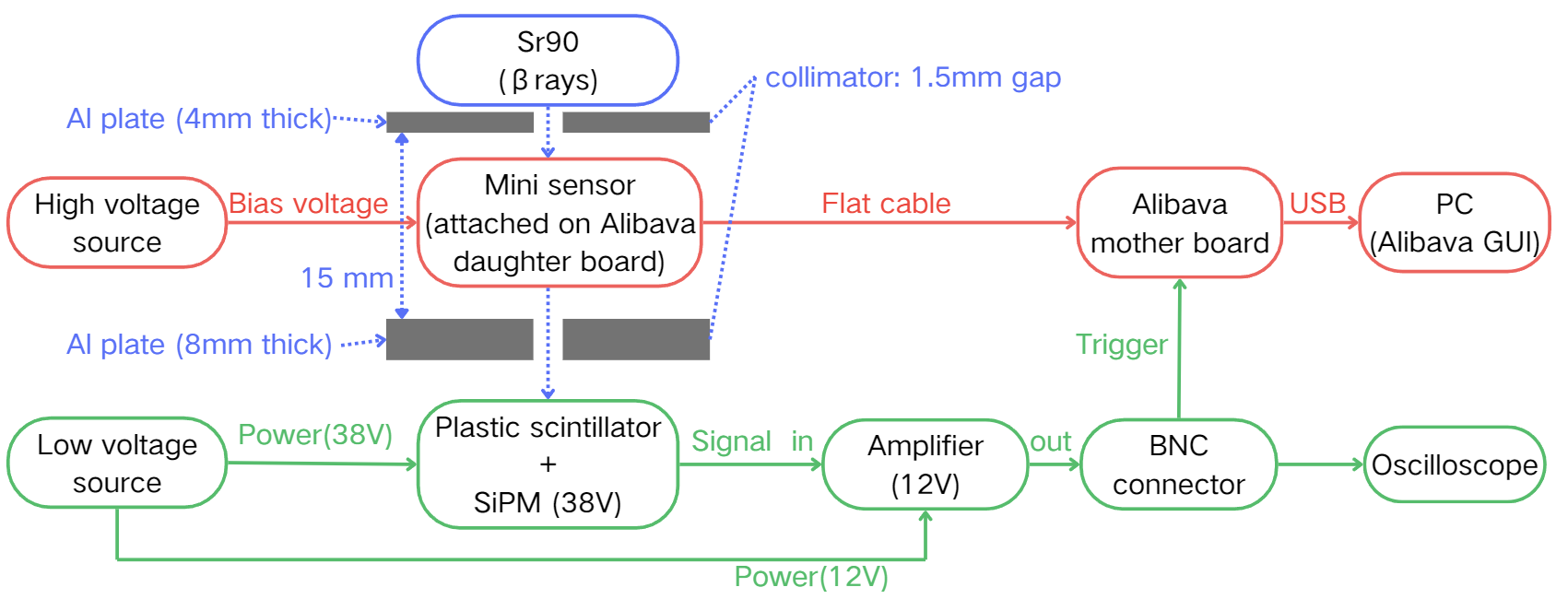}
\caption{\label{fig:2.4_CCE_connection}CCE measurement connection}
\end{figure}

The connection diagram is shown in Figure \ref{fig:2.4_CCE_connection}. It is a similar test scheme to that at other QA sites \cite{kopsalis2023establishing}.
The charge collection efficiency is measured with $\beta $-rays from a strontium-90 source. A collimator with a 1.5 mm opening below the $^{90}\text{Sr}$ source was designed to align the source with the mini sensor and exclude $\beta $-rays that cross the sensor at a significant angle. The bias voltage was applied to the mini sensor, which was mounted on the Alibava daughter board. A $\beta $-ray penetrating the mini sensor was detected by plastic scintillator and SiPM located below the mini sensor. The signal from the scintillator was amplified by the amplifier and used as the trigger signal for the Alibava system. It was also displayed on the oscilloscope, and the trigger level for the Alibava system was determined based on the waveform observed on the oscilloscope. 

\begin{figure}[!htb]
\centering
\includegraphics[width=0.6\textwidth]{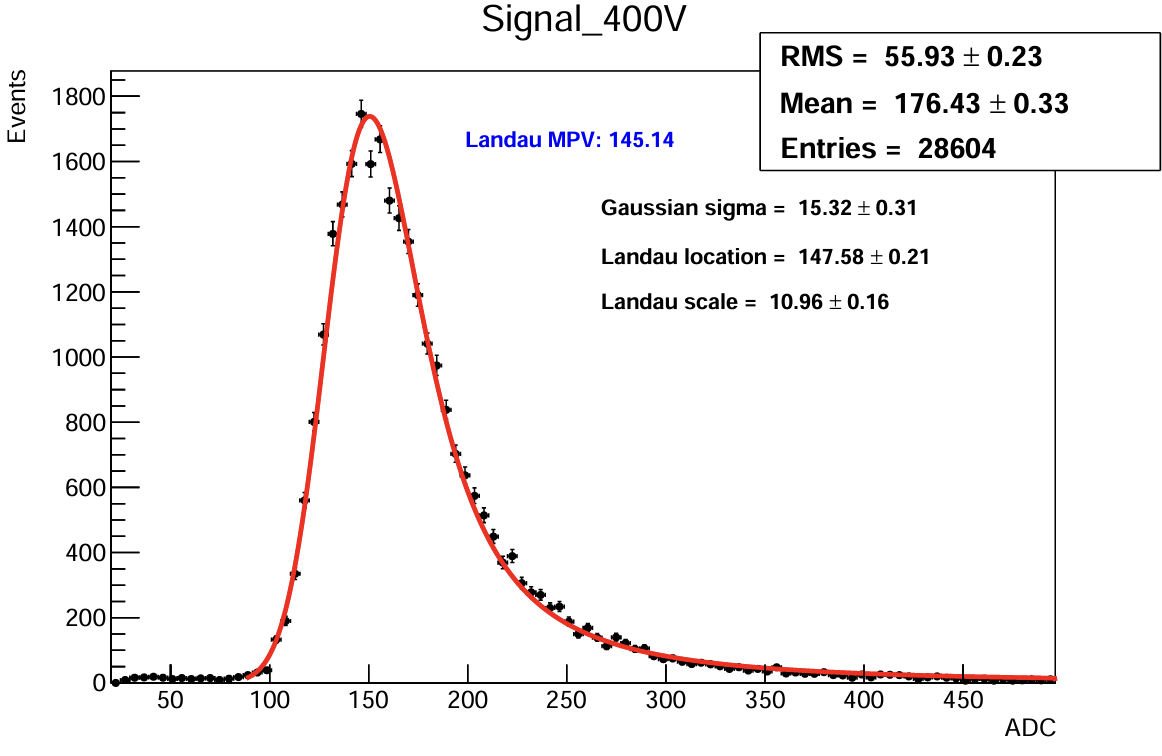}
\caption{\label{fig:2.4_CCE_unirradiated}Cluster charge distribution of un-irradiated sensor measured at 400V (black dots) fitted with a convolution of Landau and Gaussian functions (red curve).}
\end{figure}

\FloatBarrier
The collected signal was measured in 100 V steps of bias voltage from 100 V to 700 V, with 100,000 events acquired by Alibava system at each step. The event charge was determined using a clustering algorithm with a seed threshold of 3.5 times the channel’s noise level and a neighbor threshold which was 1.5 times the noise \cite{hara2020charge}. Figure \ref{fig:2.4_CCE_unirradiated} shows the distribution of cluster charges acquired by Alibava system and fitted with a convolution of Landau and Gaussian functions. The Most Probable Value of the internal Landau function (Landau MPV) can be calculated from the fitted parameters of "location" and "scale" with the formula of
$\mathrm{Landau\ MPV} = \mathrm{location} + (-0.22278 \times \mathrm{scale})$
which is confirmed to be the peak position of the Landau function independently. The Landau MPV is taken as the measure of collected charge.
\FloatBarrier

\begin{figure}[!htb]
\centering
\includegraphics[width=0.6\textwidth]{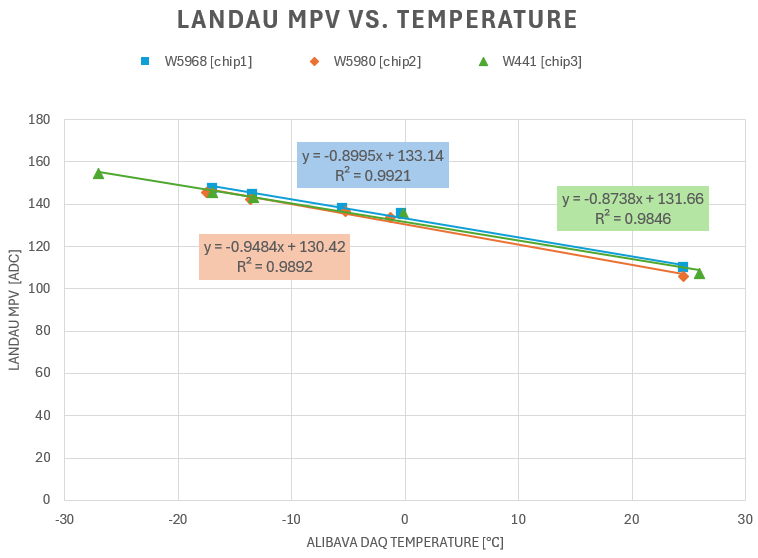}
\caption{\label{fig:2.4_CCE_calibration}Calibration of Alibava system: applied 400 V (above $V_\text{FD}$) on un-irradiated sensor for 5 different temperature points to find the relationship between Beetle chip temperature and conversion factor}
\end{figure}

The calibration of Alibava system was used to convert Landau MPV (in units of ADC) to electrons. Since the Beetle chip of the Alibava system has a temperature dependence, charge calibration relying on un-irradiated sensors of known thickness, instead of the internal calibration of the Beetle chip, enhances the reliability of charge measurements \cite{hara2016charge}. Figure \ref{fig:2.4_CCE_calibration} shows the relationship between Beetle chip temperature and conversion factor of three different Beetle chips. The gain is rather linear, but it is different for each chip. Bias voltage of 400 V (above the full depletion voltage) was applied to un-irradiated sensor at 5 different temperature points (25, 0, -5, -13, $-17\,^\circ\mathrm{C}$) to obtain the temperature-dependent calibration. The Alibava DAQ temperature, also Beetle chip temperature, was estimated from a thermistor located close to the detectors \cite{marco2008alibava}. The charge collection Q [e$^-$] above depletion voltage is assumed to be
\begin{equation}
    Q = \frac{d}{3.68} \left[ 190 + 16.3 \ln(d) \right]
\end{equation}
for the measured active thickness d (\text{$\mu$m}) of the device \cite{hara2016charge}. For the active thickness of 300 $\mu$m of the un-irradiated sensor, the charge collection Q could be around 23050 e. The conversion factor which was used to convert Landau MPV in ADC values to electrons for irradiated and un-irradiated sensors, is equal to Landau MPV (in ADC) divided by 23050 e. The collected charge calibration can be obtained for other temperatures by extrapolation of the linear dependence of the obtained conversion factors.

\FloatBarrier

\section{Results of IV, CV, and CCE measurements}
\subsection{IV measurement results}
\begin{figure}[h!]
    \centering

    \begin{subfigure}[b]{0.48\textwidth}
        \centering
        \includegraphics[width=\textwidth]{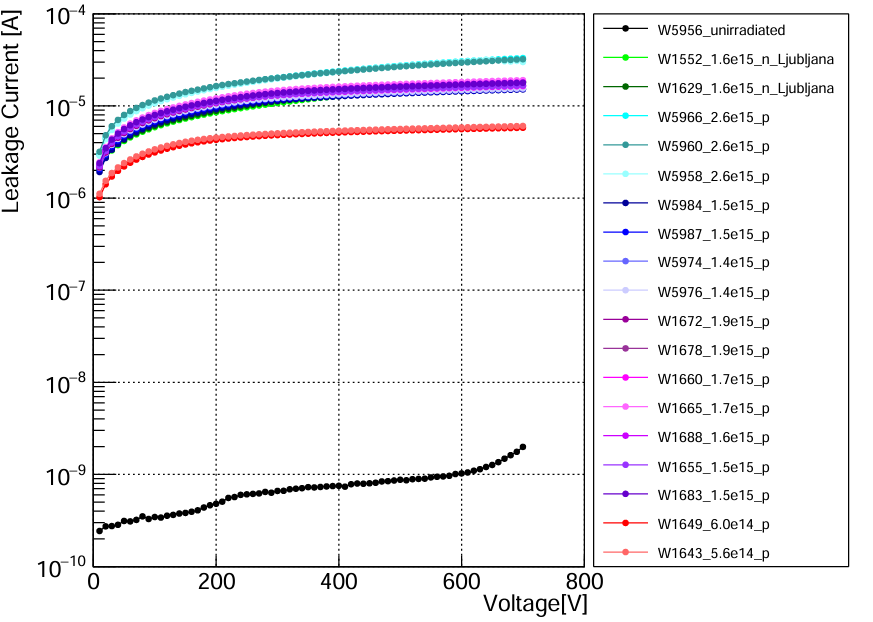}
            \end{subfigure}
    \hfill
    \begin{subfigure}[b]{0.48\textwidth}
        \centering
        \includegraphics[width=\textwidth]{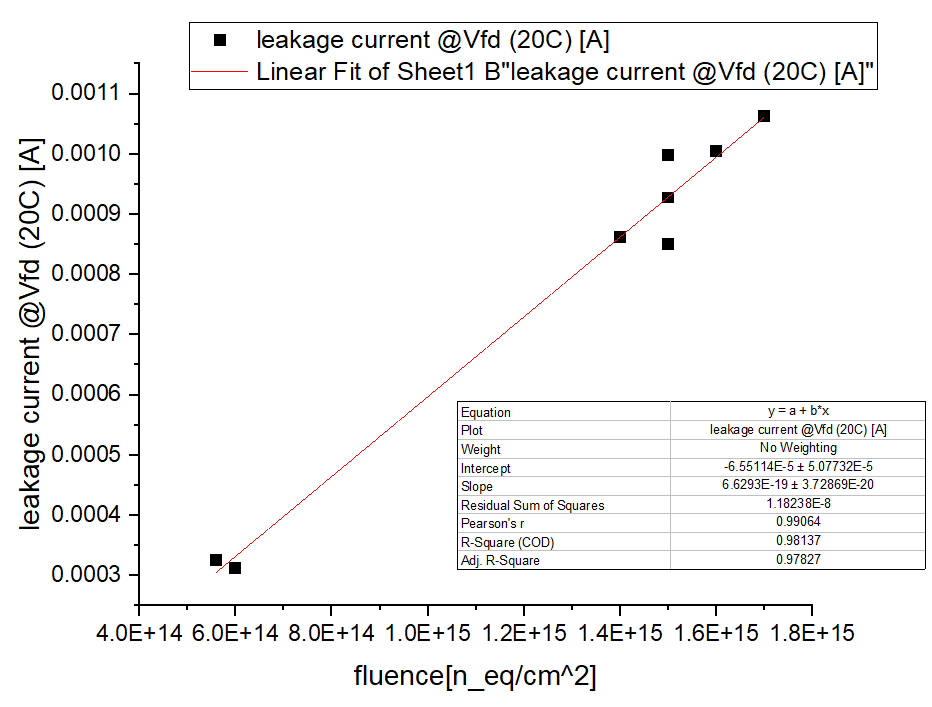}
        
    \end{subfigure}

    \caption{IV measurement results, legend: [SampleID\_ActualFluence\_IrradiationType], (all the irradiated samples were annealed at 60 $^\circ\text{C}$ for 80 min),
W1552 and W1629 shipped from Ljubljana were measured at IHEP (left). Fluence dependence of leakage current for silicon detectors (right).}
    \label{fig:3.1_leakage_current}
\end{figure}

\noindent\indent Results from the IV measurements are shown in Figure \ref{fig:3.1_leakage_current}. Tests with 18 irradiated MD8 devices were planned, however, 2 samples suffered handling issues during the mounting process. Therefore, there were no results for the two samples: one irradiated with $2.6 \times 10^{15} \, \text{n}_{\text{eq}} / \text{cm}^2$ and the other irradiated with $1.6 \times 10^{15} \, \text{n}_{\text{eq}} / \text{cm}^2$. 
The results from the other devices show that the leakage current significantly increased after proton and neutron irradiations due to bulk damage, and the leakage currents were consistent among the samples with similar fluence. The QA limit for the leakage current at full fluence is 100 $\, \mu\text{A/cm}^2$ at the bias voltage of 500 V \cite{unno2023specifications}. As shown in Figure \ref{fig:3.1_leakage_current} (left), all the measured MD8s satisfied the requirement.

\FloatBarrier

As shown in Figure \ref{fig:3.1_leakage_current} (right), for irradiated MD8, the higher the fluence, the greater the leakage current. The leakage currents of irradiated MD8s were measured at -20 $^\circ\text{C}$. Converting the leakage currents from -20 $^\circ\text{C}$ to 20 $^\circ\text{C}$ based on the formula (2) \cite{chilingarov2013currentscal}:
\begin{equation}
    I(T) \propto T^2 \exp\left(\frac{-1.21eV}{2kT}\right)
\end{equation}
and then fitting the Figure \ref{fig:3.1_leakage_current} (right) the slope ($\frac{\Delta I}{\Phi_{\text{eq}}}$) was obtained. According to the active volume ($0.76\times 0.76\times 0.03 \,\text{cm}^3$) and the formula (3) \cite{moll1999radiation}:
\begin{equation}
    \Delta I = \alpha \Phi_{\text{eq}} V
\end{equation}
the damage constant $\alpha$ is evaluated as $\left( 3.83 \pm 0.22 \right) \times 10^{-17} \, \text{A/cm}$, consistent with the value of $\alpha_{80/60} = \left( 4.06 \pm 0.04 \right) \times 10^{-17} \, \text{A/cm}$ from the literature \cite{moll1999radiation}.

\subsection{CV measurement results}
\begin{figure}[!htb]
\centering
\includegraphics[width=0.5\textwidth]{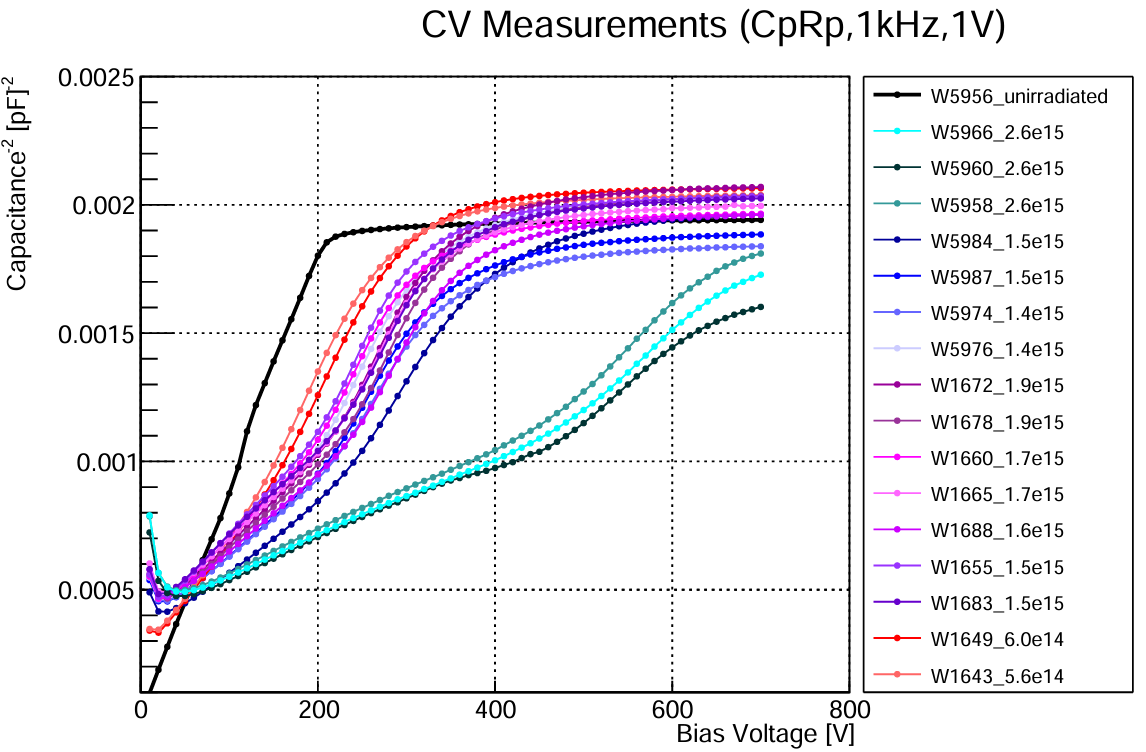}
\caption{\label{fig:3.2_CV_results}CV measurement results, legend: [SampleID\_ActualFluence] (all the irradiated samples were annealed at 60 $^\circ\text{C}$ for 80 min).}
\end{figure}

\noindent\indent Figure \ref{fig:3.2_CV_results} shows the CV measurement results from 17 MD8 pieces. The full depletion voltage ($V_{\text{FD}}$) was extracted from the CV test data. In un-irradiated detector, the $1/C^2$ vs. voltage curve is a straight line for $V < V_{\text{FD}}$ and then is a constant for $V > V_{\text{FD}}$. In irradiated detectors, $V_{\text{FD}}$ is usually determined as the intersection point of the two linear curves in the $1/C^2$ vs. voltage profile: one in the low range where the sample is under-depleted, and the other in the high voltage where over-depletion occurs, respectively \cite{CV_irrad-f-T_petterson2007}.
As shown in Figure \ref{fig:3.2_CV_results}, the full depletion voltage was around 220 V for un-irradiated MD8, meeting the QA requirement (no more than 350 V) \cite{unno2023specifications}. Evolution of the full depletion voltage as a function of fluence was clearly observed and the samples with similar fluences had consistent values.

\FloatBarrier

\subsection{CCE measurement results}
\begin{figure}[h!]
    \centering

    \begin{subfigure}[b]{0.47\textwidth}
        \centering
        \includegraphics[width=\textwidth]{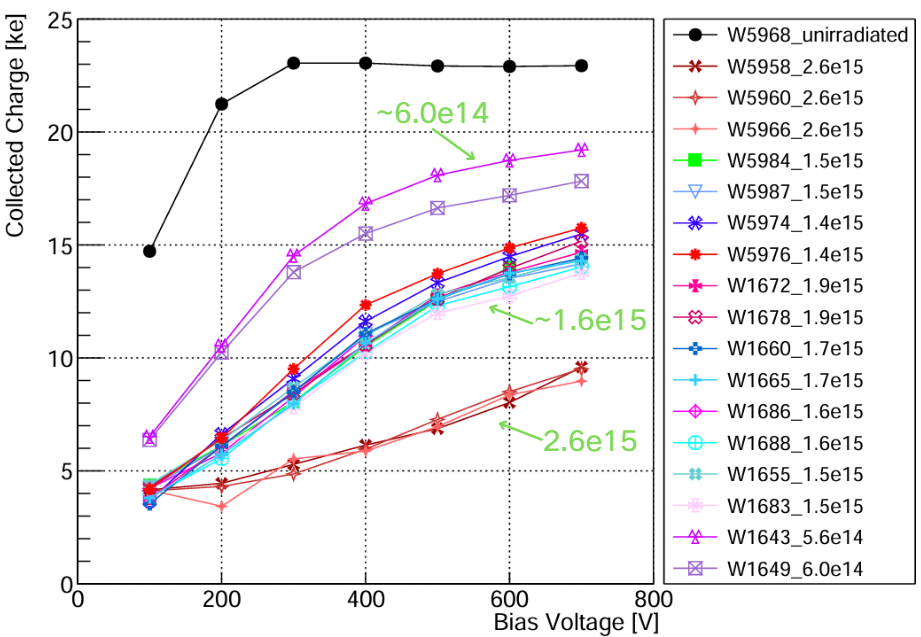}
            \end{subfigure}
    \hfill
    \begin{subfigure}[b]{0.50\textwidth}
        \centering
        \includegraphics[width=\textwidth]{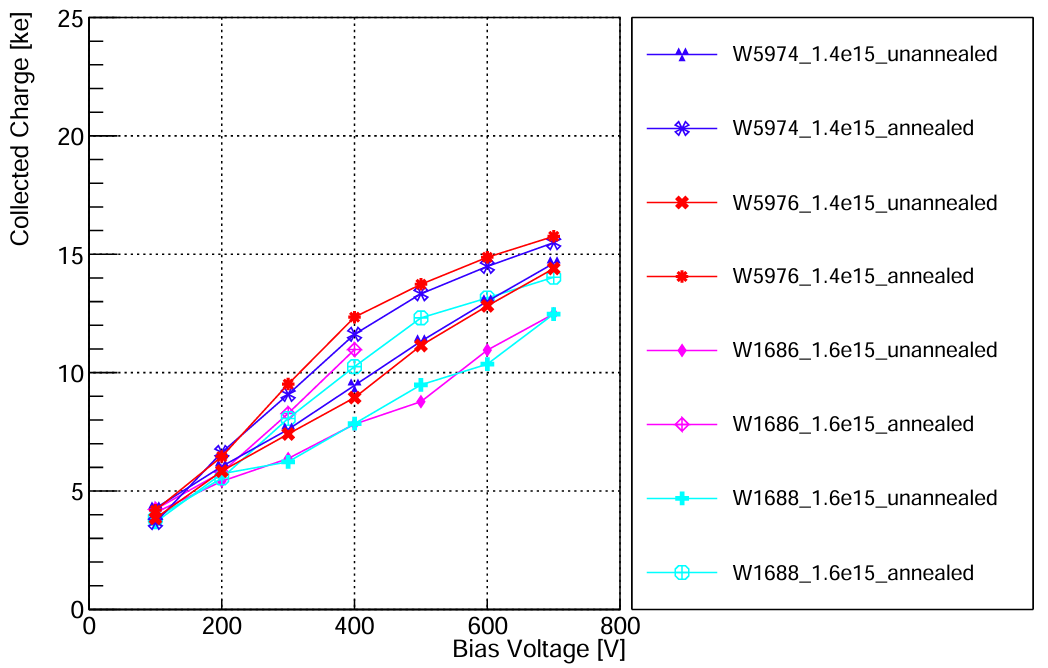}
        
    \end{subfigure}

    \caption{CCE measurement results for mini sensor (all the irradiated samples were annealed at 60 $^\circ\text{C}$ for 80 min), measured at -20 $^\circ\text{C}$, legend: [SampleID\_ActualFluence] (left). CCE measurement results before and after annealing (right).
}
    \label{fig:3.3_CCE_annealed_unannealed}
\end{figure}

\noindent\indent The collected charges from the mini sensors are shown in Figure \ref{fig:3.3_CCE_annealed_unannealed} (left). It can be seen that the collected charges are similar in groups for the sensors with similar irradiation fluences. The sensors with higher fluences yield less charges. All CCE values exceeded 6350 e threshold at 500 V, meeting the QA requirement at the end of life \cite{unno2023specifications}, except for one sensor (W1686), which broke down at 480 V.

\FloatBarrier

Four sensors irradiated at CSNS were selected to evaluate the impact of annealing. The results are shown in Figure \ref{fig:3.3_CCE_annealed_unannealed} (right). As expected, the annealed samples have larger CCE than the unannealed samples.

\begin{figure}[!htb]
\centering
\includegraphics[width=0.6\textwidth]{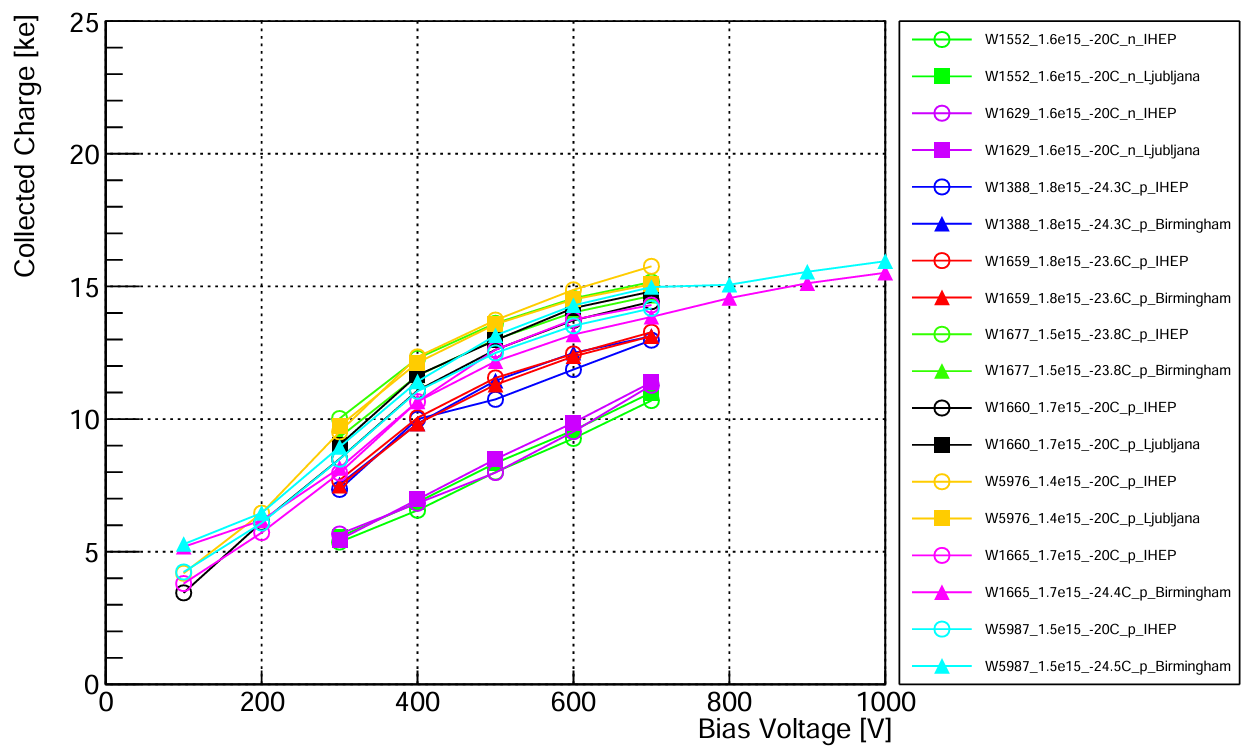}
\caption{\label{fig:3.3_CCE_crosscheck}
CCE measurement results: cross-check (all the irradiated samples were annealed at 60 $^\circ\text{C}$ for 80 min). Legend: 
 [SampleID\_ActualFluence\_Meas.Temp.\_Irrad.Type\_Meas.Site]}
\end{figure}

The irradiated samples were exchanged between the test sites: IHEP, Ljubljana, and Birmingham, to cross-check CCE measurements. The results (displayed in Figure \ref{fig:3.3_CCE_crosscheck}) showed consistency;
the CCE results measured at IHEP were in good agreement with the results from the other sites, by less than 10\%.

\FloatBarrier

\section{Conclusions}
\noindent\indent In the context of the IHEP/CSNS QA site qualification, an irradiation study was conducted using the QA test pieces from the ATLAS18 strip sensor production. A sample of 18 QA test pieces have been irradiated with 80 MeV protons provided in the APEP beamline at CSNS, at 3 fluences, up to $2.6 \times 10^{15} \, \text n_{\text{eq}} / \text{cm}^2$. The irradiation setup featured improved thermal control with graphite baseplates. The samples were subsequently tested at IHEP, including IV, CV, and CCE measurements.
The leakage currents were shown to be consistent with expectation from prior studies. The measured parameters in the group of the same fluence were shown to be consistent, indicating uniformity of the irradiation. All parameters satisfied the QA test criteria.
The irradiated samples were exchanged between IHEP, Ljubljana, and Birmingham test sites, to cross-check CCE measurements. The results measured at IHEP were consistent with those of the other sites within 10\%. The irradiation and measurement setups are available for the QA program of the ATLAS18 strip sensor production.

\section{Acknowledgements}
\noindent\indent The research was supported and financed in large part by the National Key Research and Development Program of China under Grant No. 2023YFA1605902 from the Ministry of Science and Technology. Project W2443002 supported by National Natural Science Foundation of China. This work was supported by High-performance Computing Public Platform (Shenzhen Campus) of SUN YAT-SEN UNIVERSITY; Guangdong Provincial Key Laboratory of Advanced Particle Detection Technology (2024B1212010005). The authors would like to thank the crew at the TRIGA reactor in Ljubljana for helping with irradiation. The authors acknowledge the financial support from the Slovenian Research and Innovation Agency (research core funding No. P1-0135 and project No. J1-3032). This work is part of the Spanish R\&D grant PID2021-126327OB-C22, funded by MICIU/ AEI/10.13039/501100011033 and by ERDF/EU. The work at SCIPP was supported by the US Department of Energy, grant DE-SC0010107. This work was supported by the Canada Foundation for Innovation and the Natural Sciences and Engineering Research Council of Canada.

Copyright 2024 CERN for the benefit of the ATLAS Collaboration. Reproduction of this article or parts of it is allowed as specified in the CC-BY-4.0 license.


\begin{thebibliography}{100}

\bibitem{atlasTDR}ATLAS collaboration, Technical Design Report for the ATLAS Inner Tracker Strip Detector, Tech. Rep. CERN-LHCC-2017-005, LHC/ATLAS Experiment, 2017, \DOI{10.3204/PUBDB-2017-09975}.  

\bibitem{atlasLHC} ATLAS collaboration, The ATLAS Experiment at the CERN Large Hadron Collider, J. Instrum. 3 (2008) S08003, \DOI{10.1088/1748-0221/3/08/S08003}.

\bibitem{Miguel2020QAmethodology} M. Ullán et al., Quality Assurance methodology for the ATLAS Inner Tracker strip sensor production, Nucl. Instrum. Methods Phys. Res. A 981 (2020) 164521, \DOI{10.1016/j.nima.2020.164521}.

\bibitem{hara2020charge} K. Hara et al., Charge collection study with the ATLAS ITk prototype silicon strip sensors ATLAS17LS, Nucl. Instrum. Methods Phys. Res. A 983 (2020) 164422, \DOI{10.1016/j.nima.2020.164422}.

\bibitem{LIU2022APEP} Y. Liu et al., Physical design of the APEP beam line at CSNS, Nucl. Instrum. Methods Phys. Res. A 1042 (2022) 167431, \DOI{10.1016/j.nima.2022.167431}.

\bibitem{LIHui2024169288} H. Li et al., Feasibility study of CSNS as an ATLAS ITk sensor QA irradiation site, Nucl. Instrum. Methods Phys. Res. A 1063 (2024) 169288, \DOI{10.1016/j.nima.2024.169288}.

\bibitem{LIWL2024} W. Li et al., Measurements of proton beam flux and energy of APEP using foil activation technique, Nucl. Eng. Technol. 56 (2024) 328-334, \DOI{10.1016/j.net.2023.10.004}.

\bibitem{NIEL_2000} A. Vasilescu et al., Displacement damage in silicon, on-line compilation, \url{https://rd50.web.cern.ch/rd50/NIEL/}.

\bibitem{hirose2023atlas} S. Hirose et al., ATLAS ITk strip sensor quality assurance tests and results of ATLAS18 pre-production sensors, Tech. Rep. ATL-COM-ITK-2023-001, 2023, \DOI{10.7566/JPSCP.42.011017}.

\bibitem{allport2019experimental} P. Allport et al., Experimental determination of proton hardness factors at several irradiation facilities, J. Instrum. 14 (2019) P12004, \DOI{10.1088/1748-0221/14/12/P12004}.

\bibitem{LCRe4980a_handbook} Keysight Technologies, Impedance Measurement Handbook: A Guide to Measurement Technology and Techniques, 6th Edition, \url{https://www.keysight.com/us/en/assets/7018-06840/application-notes/5950-3000.pdf}.

\bibitem{HV_capacitance2009} P. Ralston et al., High-voltage capacitance measurement system for SiC power MOSFETs, in: 2009 IEEE Energy Conversion Congress and Exposition, pages 1472–1479, IEEE, 2009, \DOI{10.1109/ECCE.2009.5316221}.

\bibitem{ullan2019quality} M. Ullán et al., Quality Assurance methodology for the ATLAS ITk Strip Sensor Production, ATL-COM-ITK-2019-053, 2019, \url{https://cds.cern.ch/record/2703676}.

\bibitem{kopsalis2023establishing} I. Kopsalis et al., Establishing the Quality Assurance programme for the strip sensor production of the ATLAS tracker upgrade including irradiation with neutrons, photons and protons to HL-LHC fluences, J. Instrum. 18 (2023) C05009, \DOI{10.1088/1748-0221/18/05/C05009}.


\bibitem{hara2016charge} K. Hara et al., Charge collection and field profile studies of heavily irradiated strip sensors for the ATLAS inner tracker upgrade, Nucl. Instrum. Methods Phys. Res. A 831 (2016) 181-188, \DOI{10.1016/j.nima.2016.04.035}.

\bibitem{marco2008alibava} R. Marco-Hernandez, A portable readout system for microstrip silicon sensors (ALIBAVA), in: Proc. 2008 IEEE Nuclear Sci. Symp. Conf. Rec., IEEE, 2008, pp. 3201-3208, \DOI{10.1109/NSSMIC.2008.4775030}.

\bibitem{unno2023specifications} Y. Unno et al., Specifications and pre-production of n+-in-p large-format strip sensors fabricated in 6-inch silicon wafers, ATLAS18, for the Inner Tracker of the ATLAS Detector for High-Luminosity Large Hadron Collider, J. Instrum. 18 (2023) T03008, \DOI{10.1088/1748-0221/18/03/T03008}.

\bibitem{chilingarov2013currentscal} A. Chilingarov, Generation current temperature scaling, PH-EP-Tech-Note-2013-001, CERN, Geneva, 2013, \url{https://cds.cern.ch/record/1511886}.

\bibitem{moll1999radiation} M. Moll, Radiation Damage in Silicon Particle Detectors: Microscopic defects and macroscopic properties, Ph.D. Thesis, Universität Hamburg, 1999, \DOI{10.3204/PUBDB-2016-02525}.

\bibitem{CV_irrad-f-T_petterson2007} M. K. Petterson et al., Charge collection and capacitance–voltage analysis in irradiated n-type magnetic Czochralski silicon detectors, Nucl. Instrum. Methods Phys. Res. A 583 (2007) 189-194, \DOI{10.1016/j.nima.2007}.






\end{thebibliography}
\end{document}